\documentstyle[preprint,prl,aps]{revtex}
\begin{document}
\preprint{TNT 94-4}
\draft

\title{Stochastic Resonance in Deterministic Chaotic Systems}

\author{A. Crisanti, M. Falcioni}
\address{Dipartimento di Fisica, Universit\`a ``La Sapienza'',
         I-00185 Roma, Italy}

\author{G. Paladin}
\address{Dipartimento di Fisica, Universit\`a dell'Aquila,
         I-67010 Coppito, L'Aquila, Italy}

\author{A. Vulpiani}
\address{Dipartimento di Fisica, Universit\`a ``La Sapienza'',
         I-00185 Roma, Italy}

\date{May 20, 1994}

\maketitle

\begin{abstract}
We propose a mechanism which produces periodic variations of the degree
of predictability in dynamical systems.
It is shown that even in the absence of noise when the control parameter
changes periodically in time, below and above the threshold
for the onset of chaos, stochastic
resonance effects appears. As a result one has an alternation of chaotic
and regular, i.e. predictable, evolutions in an almost periodic way, so
that the Lyapunov exponent is positive but  some time correlations
do not decay.
\end{abstract}


The mechanism of stochastic resonance was initially introduced as a
possible explanation of the long time climatic changes
\cite{BSV81,N82,BPSV82,BPSV83}.
In the last years it has been used in a wide class of systems in physics and
biology such as analogical circuits \cite{FH83}, ring laser \cite{McWR88},
neurology \cite{BJZMK91,CA93}, bistable systems \cite{GMMS89,Detal93},
systems with colored noise \cite{HJZM93}, see Ref.\ \onlinecite{ARW93} for a
recent review.

The phenomenon can show up in bistable systems with a periodic forcing
and a random perturbation.
A typical example \cite{BSV81,N82,BPSV82,BPSV83} is the evolution generated
by the stochastic differential equation
\begin{equation}
 \frac{dx}{dt}= - \frac{\partial V(x,t)}{\partial x} +
              \sqrt{2\sigma}\, \eta
\label{eq:1}
\end{equation}
where $V$ is a time periodic double well potential
\begin{equation}
 V(x,t)= \frac{x^4}{4} - \frac{x^2}{2} + A \, x \cos (\omega t),
\label{eq:2}
\end{equation}
and $\eta$ white noise.

It can be shown \cite{BSV81,N82,BPSV82,BPSV83} that there exists a range of
values of $A$, $\sigma$ and $\omega$ where the jumps between the two
oscillating wells are strongly synchronized, as a consequence of a sort of
resonance between periodic forcing and random perturbation.

Systems showing stochastic resonance are,
in some sense, intermediate between regular and irregular ones, since
 they are described by a random process -- the jumps do not
follow a deterministic rule -- which, nevertheless, exhibits a certain degree
of regularity. For instance, the $x$ time-correlation does not decay.

This letter shows that a similar behavior can arise in
deterministic systems close to the onset of chaos when the control parameter
varies periodically in time.
Under appropriate conditions, the time evolution shows an alternation
of regular and chaotic motion strongly
synchronized with the time variation of the control parameter.
The presence of deterministic chaos plays the role of the random
perturbation, so that it would be more correct to speak of
`chaotic resonance' rather then of stochastic resonance.
Chaotic resonance seems to be present in natural
phenomena, such as the time evolution of weather, which is
 governed by a set of non-linear  equations which surely exhibits
 deterministic chaos.
Nevertheless, one observes some elements of regularity such as high
predictability during summer, at least in the tempered regions as
the mediterranean countries, and very poor
predictability during winter.

Although the alternation of seasons cannot be described by low dimensional
systems, some qualitative features can be captured by toy
models, which can be useful as a first step toward of the comprehension
of the mechanism producing periodic variations of predictability
in short and long term climate phenomena.

We have chosen to analyze the Lorenz model \cite{L63} which is the first
geophysical dynamical system where deterministic chaos has been
 observed. We consider the original differential equations
\begin{equation}
  \left\{
         \begin{array}{ll}
              dx/dt=& 10 \, (y-x) \\
              dy/dt=& -x \, z + R(t) \, x - y \\
              dz/dt=& x \, y -{8\over 3} \, z
         \end{array}
  \right.
\label{eq:3}
\end{equation}
where the control parameter  has a periodic time variation:
\begin{equation}
  R(t)=R_0 -A \, \cos(2 \pi t/T).
\label{eq:4}
\end{equation}
The Lorenz model describes the convection of a fluid heated from below
between two layers  whose temperature difference is proportional
to the Rayleigh number $R$. In our case, the periodic variations
of $R$ roughly  mimic the seasonal changing on the solar heat inputs.

In order to get stochastic resonance effects without noise,
the average Rayleigh number $R_0$ is assumed to be close to the
threshold $R_{cr}=24.74$ for the transition
from stable fixed points to a chaotic attractor in the standard
Lorenz model. The value of the amplitude $A$
of the periodic forcing  should be such  that $R(t)$
oscillates below and above $R_{cr}$.
For very large $T$,
a good approximation of the  solution is given by
\begin{equation}
  x(t)=y(t)=\pm \sqrt{ {8\over 3} (R(t)-1)} \qquad \qquad
  z(t)=R(t)-1
\label{eq:5}
\end{equation}
which is obtained by the fixed points of the standard Lorenz model
by replacing $R$ by $R(t)$.
The stability of this solution is a rather complicated issue,
which depends on the values of $R_0$, $A$ , and $T$.
For instance, when $R_0=23.3$ and $A=4$, we numerically found
that  the solution is stable for any value of $T$,
although $R(t)$ can become larger than  $R_{cr}$.

On the other hand, it
is natural   to expect that if $R_0$ is larger than $R_{cr}$
the solution is unstable. In this case, for $A$ large enough
(at least $R_0-A < R_{cr}$)
one observes a mechanism similar to that of the stochastic resonance
in bistable systems with random forcing.
As in the case of the stochastic resonance we have a periodic variation
in the dynamics (the control parameter) and the chaos plays the role of the
noise. The value of $T$ is crucial: for large $T$ the systems behaves as
follows. It is convenient to call
\begin{equation}
    T_n \simeq nT/2 - T/4
\end{equation}
the times at which $R(t)=R_{cr}$.
For $0<t<T_1$, the control parameter  $R(t)$ is smaller than
$R_{cr}$ so that the system is stable and the trajectory
is close to one of the two solutions (\ref{eq:5}).
For $ T_1<t<T_2$, one has  $R(t)>R_{cr}$
and both solutions (\ref{eq:5}) are unstable so that the trajectory
in a short time relaxes toward a sort of `adiabatic' chaotic attractor.
The chaotic attractor smoothly changes at varying $R$
above the threshold $R_{cr}$, but if $T$ is large enough,
this dependence can be neglected in a first approximation.
However, when $R(t)$ becomes again smaller than $R_{cr}$,
the `adiabatic' attractor disappears and, in general, the system is
far from the stable solutions (\ref{eq:5}).
But, since they are attracting, the system relaxes toward them.
If the half-period $T_1$ is much larger than the relaxation time
$t_c$, in general the system follows one of the two regular solutions
(\ref{eq:5}) for $T_{2n+1} < t < T_{2n+2}$. However, there is a small but
non-zero  probability that the system has no enough time to relax to
(\ref{eq:5}) and that its evolution remains chaotic.
Figure \ref{fig:1} shows the time evolution
for $T= 300$ (a) and $T= 1600$ (b).
They provide a unambiguous numerical evidence  that
the jumps  from the  chaotic to the regular behavior
(and the contrary) are well  synchronized with $R(t)$,
with probability close to $1$
when the forcing period $T$ is very long, as in Fig.\ 1b.
On the other hand, for small value of $T$
the system often does not perform the transition
from the  chaotic to the regular behavior,  see Fig.\ 1a.

It is worth stressing that  the system is chaotic. In both cases, in fact
we found numerically that the first Lyapunov exponent is positive,
although  the correlation function
of the variable $z$ does not decay. This is due to the presence of
strong correlation between the regular intervals.

Figure \ref{fig:2} shows the probability distribution of the lengths of the
irregular interval. One observes peaks around $T/2$, $ 3T/2$,
$5T/2 \cdots  $, while the envelope of the probability distribution
decreases exponentially. This feature can be easily explained.
 At $t=T_{2n} \,\,\,\, (n=1,2 \cdots)$
the system will be in some part of the `adiabatic' chaotic
attractor. The phase space is divided
into two regions $\Omega_1$ and $\Omega_2$ such that if
${\bf x}(T_{2n})$ is contained in $\Omega_1$ the trajectory during
the following half-period will be very close to one of the two solutions
(\ref{eq:5}).
On the other hand the points ${\bf x}(T_{2n})$ contained in $\Omega_2$
generate trajectories which  remain far from  (\ref{eq:5}).
Calling $\pi$ the measure of the region  $\Omega_2$ and noting
that in the irregular intervals the correlations decay very fast,
it follows that the probability, $P_n$, that the lengths of the irregular
interval is close to $T_{2n+1}$ is $P_n\simeq \widetilde{p} ^n=\exp(-c\,n)$
with $c=-\ln \widetilde{p}$.

This feature has been observed in many other systems
exhibiting stochastic resonance \cite{BJZMK91,CA93,ZMJ90,LBM91,ILD93}.

The probability of jumping a regular interval, $\pi$,
decreases with the period of the forcing $T$, of course.
Figure\ \ref{fig:3} shows that in the Lorenz model (\ref{eq:3}), the
probability $P(T)$ to have an irregular interval longer than $T$ decreases as:
\begin{equation}
 P(T)=  \int_{T}^{\infty}p(\tau)\, d\tau \simeq e^{-\alpha \, T}
\label{eq:6}
\end{equation}
where $p(\tau)$ is the probability distribution of the length of the
irregular interval.

Without entering in the details, we briefly discuss the effect of
a random forcing, of strength $\sigma$, in the case where
 $R(t)-R_{cr}$  changes sign during the time evolution but the
solutions (\ref{eq:5}), in the absence of the noise, are stable.
In practice, we consider the Langevin equation
\begin{equation}
  \left\{
       \begin{array}{ll}
              dx/dt=& 10 \, (y-x) + \sqrt{2\sigma}\, \eta_1 \\
              dy/dt=& -x \, z + R(t) \, x - y +
                                   \sqrt{2\sigma}\, \eta_2 \\
              dz/dt=& x \, y -{8\over 3} \, z +
                                   \sqrt{2\sigma}\, \eta_3
       \end{array}
    \right.
\label{eq:7}
\end{equation}
where $\eta_i(t)$ are uncorrelated white noises i.e.
$<\eta_i(t)\eta_j(t')>=\delta_{ij} \delta(t-t')$.

The numerical study of the model (\ref{eq:7}) reveals a
phenomenology very close to the original stochastic resonance
\cite{BSV81,N82,BPSV82,BPSV83}. For small values of $\sigma$ one has
the same qualitative behavior obtained
at $\sigma=0$, while for $\sigma$ slightly larger than
a critical value $\sigma_{cr}$ one has an alternation of regular
and irregular motions. Now  the Lyapunov
exponent, computed treating the noise as an usual time-dependent term,
is negative, i.e. two trajectories, initially close, with the
same realization of the random forcing do not separate
but stick exponentially fast. We stress that the Lyapunov exponent computed
in the above method is neither unique nor the most physically relevant
characterization of the complexity of noisy systems \cite{marsici}.

It is not difficult to give a rough argument for the above features.
In the time interval where $R(t)<R_{cr}$, because of the random noise,
the distance $\delta$ between the state of the system ${\bf x}$
and the solutions (\ref{eq:5}) is $O(\sqrt{\sigma})$. During the half-period
$T_{2n+1}<t<T_{2n+2}$, the typical distance $\delta$ grows exponentially:
\begin{equation}
  \delta(t)\sim \sqrt{\sigma} e^{ c (t-T_{2n+1})}
\label{eq:8}
\end{equation}
Very roughly,  $c$ is related to the largest real part of the
eigenvalues of the stability matrix computed along the solutions (\ref{eq:5}).
Calling $L$ the size of the `adiabatic attractor', if the strength of
the random forcing is large enough, i.e.
\begin{equation}
  \sigma > \sigma_{cr} \sim L^2 e^{-c \, T}
\label{eq:9}
\end{equation}
the system can  jump into  the `adiabatic attractor'
at a time between $T_{2n+1}$ and $T_{2n+2}$  and one has the same
behavior shown in Fig.~1b.

This feature is quite similar to the original stochastic resonance, as
the central role is played by the forcing term.
Let us stress that the critical value $\sigma_{cr}$ decreases
very quickly with the period $T$.

In conclusion we have shown that the phenomenology of the stochastic
resonance can appear in a dynamical system even in the absence of a random
perturbation, when there is a  periodic time variation of the control
parameter around the onset of chaos.
 Instead of the two minima in the double  well potential
 considered by the original stochastic resonance,
 one  has two dynamical states
of the system: chaotic and regular. The role of the noise
is played by the chaotic evolution itself.
 It is worth noting that one needs that
 the period $T$ of the control parameter variations
 should be much larger than the internal
relaxation time $t_c$ toward the regular solution
 of the unperturbed system.

Stochastic resonance in
chaotic systems has relevant consequence on the predictability problem.
It shows that the predictability time is not trivially related to the
Lyapunov exponent if $T$ is large enough. During the regular intervals,  one
has an almost perfect predictability while in the irregular intervals the
predictability time is given by the inverse of the Lyapunov exponent.
Moreover, we have shown that there is a non-zero probability
(vanishing when $T \to \infty$) to skip a regular interval.
Using a pictorial language, we could say
that  the regular interval corresponds to the summer evolution,
 while the irregular one to winter.
 Although the Lorenz model is too naive for any attempt of a realistic
description, it allows us to reproduce some important features
 of weather forecasting which motivated our work:
 the forecasting is limited up to a time proportional to
the inverse Lyapunov exponent of the system during winter;
 there is a very high predictability in summer;
 there is a small but not negligible probability to
 have very bad summers (jumps of the regular intervals)
where the weather is unpredictable.

\acknowledgments
We are grateful to M. Serva for many useful discussions.
MF, GP and AV acknowledge the financial support of the INFN
 through the {\it Iniziativa specifica FI3}.

\begin{figure}
\caption{Model with  $A=4$, $R_0=25.5$.
         $z$ as a function of $t/T$ for $T=300$ (1a) and $T=1600$ (1b).
        }
\label{fig:1}
\end{figure}

\begin{figure}
\caption{Model with  $A=4$, $R_0=25.5$.
         Probability density, $p$, to have an irregular interval
         $\tau =\Delta t /T$, for $T=300$.
        }
\label{fig:2}
\end{figure}

\begin{figure}
\caption{Model with  $A=4$, $R_0=25.5$.
         Probability of jumping one  or more regular intervals, $P$, as a
         function of the forcing period, $T$.
        }
\label{fig:3}
\end{figure}

\end{document}